\documentclass[useAMS,usenatbib]{mn2e}

\usepackage[dvipdfm]{graphicx}
\def\lya{{\rm\,Ly$\alpha$}}
\def\ha{{\rm\,H$\alpha$}}

\def\oii{{\rm\,[O{\sc ii}]}}

\def\oiii{{\rm\,[O{\sc iii}]}}

\def\pa{{\rm\,Pa$\alpha$}}
\def\msun{{\rm M}$_{\odot}$}
\def\mn{\ifmmode {\mu {\rm m} ~}\else {$\mu $m ~}\fi}

\title[Identification of the progenitors of rich clusters and member galaxies in rapid formation at z$>$2]{Identification of the progenitors of rich clusters and member galaxies in rapid formation at z$>$2}
\author[R. Shimakawa, T. Kodama, K.-i. Tadaki, I. Tanaka, M. Hayashi, and Y. Koyama]{Rhythm Shimakawa$^{1,2}$\thanks{E-mail:
rhythm@naoj.org (RS); t.kodama@nao.ac.jp (TK)}, Tadayuki Kodama$^{2,3}$\footnotemark[1], Ken-ichi Tadaki$^{3}$, Ichi Tanaka$^{1}$,
 \newauthor
 Masao Hayashi$^{4}$, and Yusei Koyama$^{3}$ \\
$^{1}$Subaru Telescope, National Astronomical Observatory of Japan, 650 North A$\textquoteright$ohoku Place, Hilo, HI 96720, USA \\
$^{2}$Department of Astronomy, School of Science, Graduate University for Advanced Studies, Mitaka, Tokyo 181-8588, Japan \\
$^{3}$Optical and Infrared Astronomy Division, National Astronomical Observatory, Mitaka, Tokyo 181-8588, Japan \\
$^{4}$Institute for Cosmic Ray Research, The University of Tokyo, Kashiwa, Chiba 277-8582, Japan}
\begin{document}

\date{Accepted 1988 December 15. Received 1988 December 14; in original form 2013 August 16}

\pagerange{\pageref{firstpage}--\pageref{lastpage}} \pubyear{2014}

\maketitle

\label{firstpage}

\begin{abstract}
We present the results of near-infrared spectroscopy of \ha\ emitters (HAEs) associated with 
two protoclusters around radio galaxies (PKS1138-262 at $z$=2.2 and USS1558-003 at 
$z$=2.5) with Multi-Object Infrared Camera and Spectrograph (MOIRCS) on the Subaru telescope. 
Among the HAE candidates constructed from our narrow-band imaging, we have 
confirmed membership of 27 and 36 HAEs for the respective protoclusters, with a success 
rate of 70 per cent of our observed targets. 
The large number of spectroscopically confirmed members per cluster has enabled us for 
the first time to reveal the detailed kinematical structures of the protoclusters at $z$$>$2.
The clusters show prominent substructures such as clumps, filaments and 
velocity gradients, suggesting that they are still in the midst of rapid construction to grow 
to rich clusters at later times. 
We also estimate dynamical masses of the clusters and substructures assuming
their local virialization.
The inferred masses ($\sim$10$^{14}$M$_\odot$) of the protocluster cores are consistent with 
being typical progenitors of the present-day most massive class of galaxy clusters 
($\sim$10$^{15}$M$_\odot$) if we take into account the typical mass growth history of clusters.
We then calculated the integrated star formation rates of the protocluster cores
normalized by the dynamical masses, and compare these with lower redshift descendants.
We see a marked increase of star-forming activities in the cluster cores, by almost three orders 
of magnitude, as we go back in time to 11 billion years ago; this scales as (1$+$$z$)$^6$.
\end{abstract}


\section{Introduction}

In protoclusters at $z$$>$2, characteristic relations seen in low-$z$ clusters ($z$$<$1) such as 
the colour-magnitude relation, break down as galaxies enter into their formation phase \citep{Kodama:2007}. 
Since the galaxies in those protoclusters are destined to evolve into early-type galaxies
in rich clusters today, the protoclusters provide us with unique laboratories in which to investigate 
directly the formation mechanisms of early-type galaxies and their environmental dependence, 
through comparison with field galaxies at similar redshifts. 
It is therefore essential to investigate characteristics of protoclusters at $z$$>$2 systematically, 
in order to know how the star forming (SF) activities in high density regions at
high $z$ are intrinsically biased, and how they are affected externally by their surrounding
environments to establish the strong environmental dependence seen in the present-day Universe.

With this motivations, we have been conducting a systematic study of protoclusters at
$z$$>$1.5 with Subaru, as the project `Mahalo-Subaru' ({\it Mapping HAlpha and Lines
of Oxygen with Subaru}; for more detail see \citealt{Kodama:2013}). 
We have conducted narrow-band (NB) imaging with many customized NB filters, and 
have successfully identified \ha\ or \oii\ emitters candidates that are physically 
associated with the protoclusters.
The following two objects are among the richest systems ever identified: 
USS1558--003 at $z$$=$2.53 (\citealt{Hayashi:2012}, hereafter H12), and
PKS1138--262 at $z$$=$2.16 (\citealt{Koyama:2013}, hereafter K13).
Since they both show large excesses in number densities of SF galaxies,
these protoclusters are probably still in the vigorous formation process.
Our observations have revealed high SF activities towards the cores of protoclusters
at $z$$>$2. 
The peak of SF activity traced by the line emitters is shifted from dense cluster cores
to lower density outskirts and filamentary outer structures with time from $z$$\sim$2.5 to $z$$\sim$0.4,
indicating the inside-out growth of clusters.

These results are all intriguing but we need to confirm them and investigate the physical
properties of protocluster galaxies in much greater detail with spectroscopic follow-up
observations.
This Letter reports the first results of our near-infrared (NIR) spectroscopies of HAEs
in the two richest protoclusters at $z$$>$2.
We first describe our targets and the method of spectroscopic observations and then present
the kinematical structures of spectroscopically confirmed members.
We also discuss SF activities in these two systems through comparison with lower redshift
counterparts.
We assume a $\Lambda$-dominated cosmology with of $\Omega_M$=0.3, 
$\Omega_\Lambda$=0.7 and $H_0$=70 km s$^{-1}$ Mpc$^{-1}$. 

\section{Observation \& Data Reduction}

Our targets are selected on the basis of NB \ha\ imaging together with broad-band imaging
of the two protoclusters around the radio galaxies (RGs), namely PKS1138--262 ($z$=2.16)
and USS1558--003 ($z$=2.53) (hereafter PKS1138 and USS1558, respectively).
First we sample line emitters that show excess fluxes in the NB and then use a broad-band 
colour-colour diagram ($BzK$ or $rJK$) to separate \ha\ emitters at the
cluster redshift from contaminant \oiii/\oii/\pa\ emitters at other redshifts
(see K13 and H12 for more details). 
We select the emitters with \ha\ fluxes larger than $2.5$$\times$$10^{-17}$ erg s$^{-1}$ cm$^{-2}$
as estimated from the NB imaging (K13; H12), which corresponds to
SFR$\sim$19--25 M$_\odot$ yr$^{-1}$ for PKS1138 and USS1558 respectively,
using the \citet{Kennicutt:1998} conversion. 
We have identified 48 and 68 HAEs candidates in the vicinity of
PKS1138 and USS1558 respectively (K13; H12).
We used {\it MOIRCS}, a NIR imager and spectrograph \citep{Ichikawa:2006, Suzuki:2008} 
mounted on the 8.2-m Subaru Telescope on Mauna Kea.
It provides a multi-object spectroscopic (MOS) capability
at $0.9$$\sim$$2.5$ $\mathrm{\mu m}$ with a $4'$$\times$$7'$ filed of view
covered by two {\it HAWAII-2} $2048$$\times$$2048$ arrays with the spatial resolution 
of $0.117$ arcsec/pixel. 
We used a low-resolution grism ({\it HK500}: R$\sim$500 for 0.8 arcsec slit width) for 5 masks, 
and a high resolution grism ({\it VPH-K}: R$\sim$1700 for 0.8 arcsec slit width; see 
\citealt{Ebizuka:2011}) for one of the 3 masks for PKS1138. 
We made more than 15 slits per mask.
In total, 98 objects were observed (some targets were redundant). 

We spent 5 nights (March--April in 2013) under 0.6--1 arcsec seeing conditions for most of the 
time; however, one night was completely lost due to bad weather.
The net integration times were longer than 2 h per mask. 
A summary of the observations is given in Table \ref{tab1}. 

Data reduction was carried out using MOIRCS MOS Data Pipeline ({\sc mcsmdp}\footnotemark[1], 
\citealt{Yoshikawa:2010}) which is a data reduction package based on {\sc iraf}\footnotemark[2] 
for spectroscopic data of MOIRCS. 
This pipeline executes bad pixel masking, cosmic-ray rejection, pairwise frame subtraction, 
correction for distortion, and combination of positive and negative spectra semi-automatically. 
A wavelength calibration is performed with OH lines and a flux calibration is performed
using standard stars with a spectral type of A0V. 
For USS1558, we checked the result of wavelength calibration carefully by comparing it 
with thorium argon lines, since \ha\ lines are located beyond $\lambda$$=2.3$ $\mathrm{\mu m}$ 
where there is no remarkable OH line.
However, it should be noted that since thorium argon lines are not strong
enough at the long wavelength range of 2.3--2.5 $\mathrm{\mu m}$, we have 
larger uncertainties for the USS1558 members. 
Therefore we use an \oiii\ emission line, if available, to estimate the redshift of a USS1558 member.
The root-mean-square (RMS) of wavelength calibration is about $\pm$3$\mathrm{\AA}$, which 
corresponds to $\pm$40--50 km s$^{-1}$. 
This error is negligible for determination of the velocity dispersion of each cluster.

\begin{table}
\centering
 \caption{Summary of the observations. Columns: 
 (1) cluster name, (2) grism name, (3) resolution with 0.8 arcsec slit width, 
 (4) integration time and (5) the number of observed targets.
 HK500 and VPH-K indicate low- and high-resolution grisms covering the
 ranges of 1.3--2.5 and 1.9--2.3 $\mu m$, respectively.}
 \begin{tabular}{@{}ccccc@{}}
 \hline
  Cluster Name & Grism & R & Integ. time & Targets \\
  (1) & (2) & (3) & (4) & (5) \\
\hline
PKS1138-262 & HK500 & 513 & 120 min & 23 \\
(11:40:48,$-$26:29:08) & HK500 & 513 & 161 min & 19 \\
$z$$=$2.16 & VPH-K & 1675 & 225 min & 18 \\
\hline
USS1558-003 & HK500 & 513 & 180 min & 25 \\
(16:01:17,$+$00:28:48) & HK500 & 513 & 276 min & 19 \\
$z$$=$2.53 & HK500 & 513 & 175 min & 15 \\
\hline
\end{tabular}
\label{tab1}
\end{table}

\begin{figure}
\centering
\includegraphics[width=80mm]{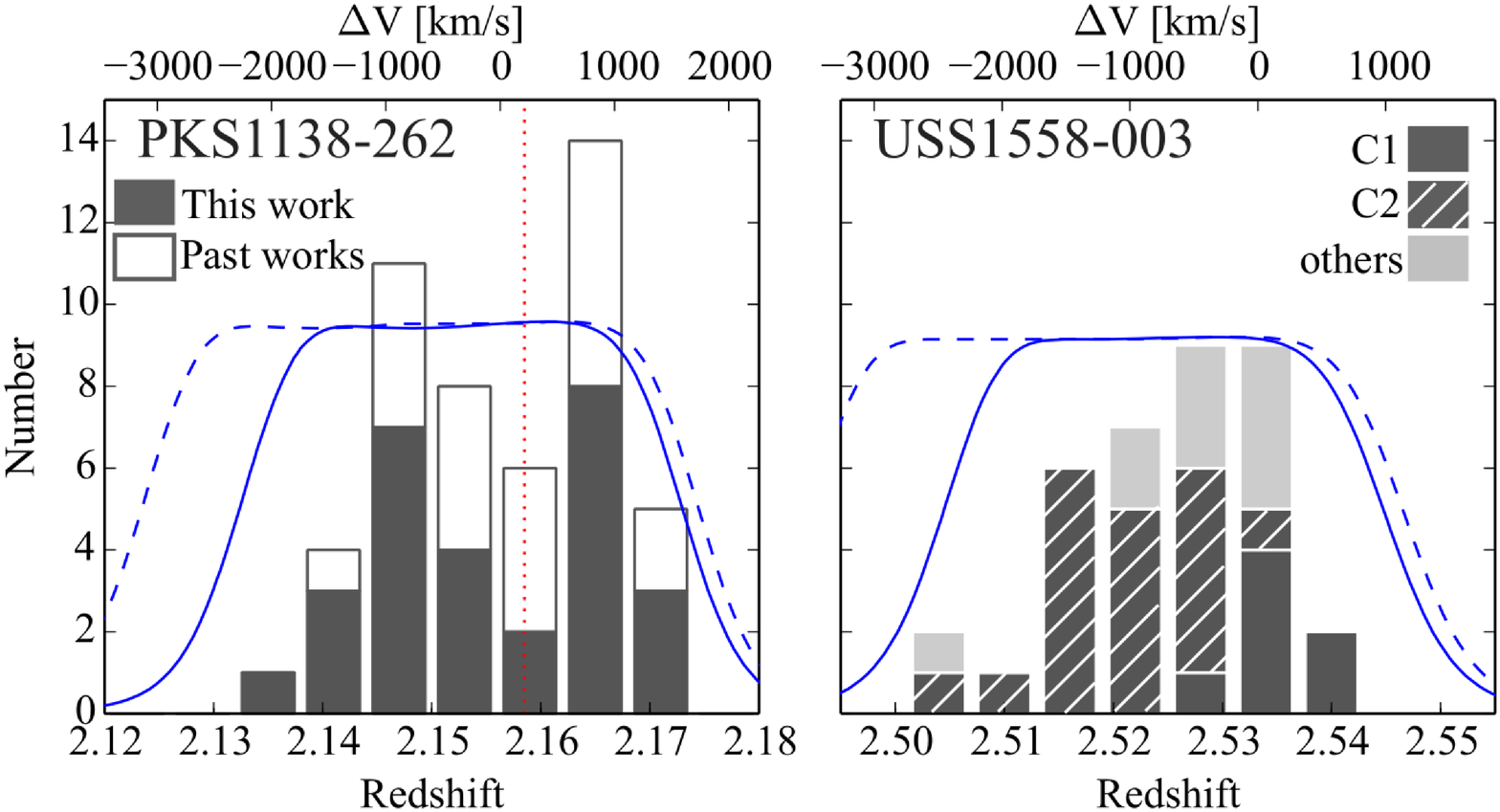}
 \caption{
Distribution of redshifts and peculiar velocities with respect to the RGs 
in PKS1138 (left) and USS1558 clusters (right).
Blue solid curves show the response curves of the NB filters used in K13 and H12, 
namely NB2071 ($\lambda_c$$=$2.070 $\mathrm{\mu m}$, FWHM$=$0.027 $\mathrm{\mu m}$) 
and NB2315 ($\lambda_c$$=$2.317 $\mathrm{\mu m}$, FWHM$=$0.026 $\mathrm{\mu m}$) 
respectively, and both of the two filters belong to T.\ Kodama. 
Blue dashed curves represent the range of wavelength shifts of the transmission curves within 
the FoV$^3$.
Left: black and white histograms present our results (including the RG) and those of 
past works \citep{Kurk:2000, Kurk:2004, Croft:2005, Doherty:2010}. 
The red dotted line indicates the location of a strong OH line at $\lambda$=2.073 nm. 
Right: confirmed cluster members including the RG are plotted. 
Black solid, slashed, and grey solid histogram show the galaxies in clumps 1, 2 and 
outside of them in the protocluster, respectively.
The bin size is $\Delta$$z$=0.006 in both panels. 
The top axis shows peculiar velocities ($\Delta$V) with respect to each RG. }
\label{fig1}
\end{figure}

\footnotetext[1]{Available at www.naoj.org/Observing/DataReduction/}
\footnotetext[2]{IRAF is distributed by National Optical Astronomy Observatory and available at iraf.noao.edu/}
\footnotetext[3]{subarutelescope.org/Observing/Instruments/MOIRCS/}

\section{Results}

\begin{figure*}
\centering
\includegraphics[width=170mm]{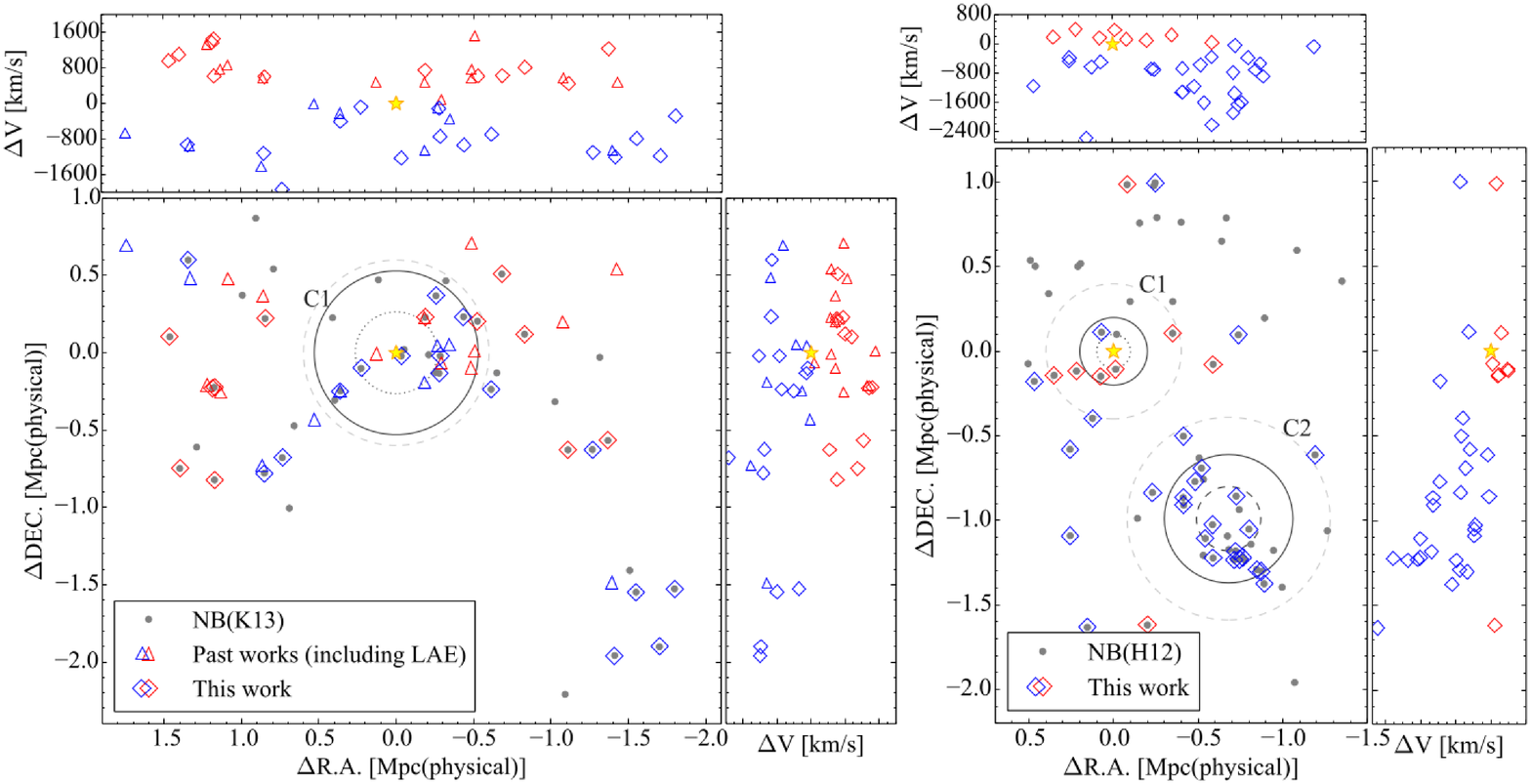}
 \caption{The kinematical structures of HAEs in (1) PKS1138 (left) and (b) USS1558 clusters (right). 
Grey dots represent HAE candidates detected in our NB imaging (K13 and H12). 
Diamonds show the members newly confirmed in this study. Triangles are the confirmed \ha\ and \lya\ 
emitters (LAEs) in the previous works \citep{Kurk:2000, Kurk:2004, Croft:2005, Doherty:2010}. 
Blue and red symbols are separated by blue- and redshifted galaxies, respectively, relative to the RGs (star mark). 
We identify three groups, namely PKS1138-C1, USS1558-C1, and USS1558-C2, and they are shown by grey dashed circles. 
Solid and dotted black circles indicate R$_{200}$ and 0.5$\times$R$_{200}$, respectively.}
\label{fig2}
\end{figure*}

We first search for emission lines in the wavelength range of NB filters, where \ha\ line candidates
have been already detected by the past NB imaging surveys. 
Next they are fitted by Gaussian curves with {\sc specfit}
\citep{Kriss:1994} which is distributed within {\sc stsdas}\footnotemark[4]
layered on top of the {\sc iraf} environment.
We normally apply a single Gaussian fitting, but sometimes apply a multi-Gaussian 
fitting for a broad or multiple emission line, and the chi-square minimization technique 
is used to best fit the line profile.
Further details of the line fitting will be presented in our forthcoming full Paper I\hspace{-.1em}I (in preparation).

\footnotetext[4]{Available at www.stsci.edu/institute/software\_hardware/stsdas/}

\begin{table}
\centering
 \caption{Summary of spectroscopic confirmations. Columns: 
 (1) cluster name as shown in Fig. \ref{fig2}, (2) newly confirmed number of members, (3) mean peculiar 
velocity, (4) velocity dispersion, (5) R$_{200}$ and (6) cluster mass (M$_{cl}$) calculated from
\citet{Finn:2005}.}
 \begin{tabular}{@{}cccccc@{}}
 \hline
  Cluster & New & $\langle$$\Delta$V$\rangle$ & $\sigma_{cl}$ & R$_{200}$ & M$_{cl}$ \\
  name & members & [km/s] & [km/s] & [Mpc] & [10$^{14}$M$_\odot$] \\
  (1) & (2) & (3) & (4) & (5) & (6) \\
\hline
PKS1138 all  & 27 & $-$41 & 882\footnotemark[5] & --- & --- \\
PKS1138 C1  & 9 & $+$9 & 683\footnotemark[5] & 0.53\footnotemark[5] & 1.71\footnotemark[5] \\
 \hline
USS1558 all  & 36 & $-$717 & 756 & --- & --- \\
USS1558 C1  & 7  & $+$121 & 284 & 0.19 & 0.10 \\
USS1558 C2  & 19 & $-$1042 & 574 & 0.38 & 0.87 \\
\hline
\end{tabular}
\label{tab2}
\end{table}
\footnotetext[5]{estimated including \lya\ emitters in the literature}

In the obtained spectra, we identified one or more new emission lines at above 3$\sigma$ levels
for 27 and 36 SF galaxies in PKS1138 and USS1558 clusters, respectively. 
Here, 1$\sigma$ threshold is defined as a standard deviation of flux densities around
each emission line, excluding the line itself.
The limiting magnitude of this spectroscopy (1$\sigma$) is $m_{AB}$=22.2--22.6 in the K-band. 
The completeness of our observation is 76 per cent (90 observed out of 116 candidates) and the efficiency
or success rate is 70 per cent (63 confirmed out of 90 observed).
A summary of our spectroscopic confirmations is presented in Table \ref{tab2}.
We note that velocity dispersions and R$_{200}$ are measured including the cluster members reported
in the past studies. 
The histograms (Fig. \ref{fig1}) show the redshift distributions of HAEs in the two protoclusters. 
For PKS1138, we also plot cluster members identified by past works \citep{Kurk:2000, Kurk:2004, 
Croft:2005, Doherty:2010}, and in total 49 members have been spectroscopically confirmed, including the RG.

It should be noted that a relatively strong OH sky line is located at $\lambda$=2.073 nm,
and it significantly affects our line detectability at the specific redshift interval of
$z$=2.156--2.162 for the \ha\ emission line.
This actually contributes to the dip seen in the redshift distribution at the corresponding bin.
The velocity dispersions of `PKS1138' all and `PKS1138 C1' shown in Table \ref{tab2}
may decrease by 20--30 km s$^{-1}$ if we take this effect into account.

As seen in Fig. \ref{fig1}, cluster members are mostly located at redshifts where the response
curve has the maximum sensitivity. 
The large numbers of newly confirmed members in both clusters have confirmed that these systems
are indeed rich protoclusters hosting lots of highly SF young galaxies with typical
star formation rates (SFRs) of $\sim$15--800 \msun yr$^{-1}$ after dust extinction correction. 
Since they are wide spread and show clear substructures as shown in Fig.\ \ref{fig2}, 
the protoclusters are right in the phase of vigorous assembly.
It should be noted, however, that only 15 and 21 galaxies out of 27 and 36 emitters
in PKS1138 and USS1558, respectively, are confirmed with detections of more than one
emission lines.
Although our colour selections works well to discriminate \ha\ line at the cluster redshifts from 
other contaminant lines at wrong redshifts (K13; H12), we cannot fully rule out some 
contamination from the fore- or background of the protoclusters. 
In fact, we identified one background \oiii\ emitter with doublet lines in each field. 
Therefore, among those 38 targets with redshifts determined with multiple lines, 
36 were identified as HAEs at the cluster redshifts.
Contamination level by fore- or background galaxies would then be $\sim$5 per cent,
and thus has little impact on the current study.

\section{Discussion and Conclusions}

Kinematical structures of distant protoclusters provide essential information
on the mass assembly history of galaxy clusters. 
Fig. \ref{fig2} presents the spatial and redshift (or radial velocity)
distributions of the HAEs (and 15 confirmed LAEs in the case of PKS1138). 
The blue and red symbols separate the members according to their redshifts or
radial velocities: those approaching or receding, respectively, with respect to the RGs.

From the velocity dispersion, we now estimate R$_{200}$ of each protocluster, which is 
the radius within which the averaged matter density is 200 times larger than the critical density. 
The dynamical mass of each system (M$_{cl}$) is also measured using the virial theorem 
\citep{Finn:2005}, assuming local virialization.

\subsection{PKS 1138-262 ($\mathbf{z}$=2.16)}
PKS1138 is among the most familiar protocluster at $z$$\sim$2,
and has been intensively studied by many researchers
(e.g. \citealt{Kurk:2000}). 

We see some velocity structures across the protocluster (Fig. \ref{fig2}a)
as well as the spatial structure.
The galaxies near the RG within the dashed circle denoted as C1 seem to be 
relatively well mixed in velocity. 
However, the outer region are more structured. 
The most prominent structure is a linear filament extending towards south-east
direction from the RG.
This almost perfectly aligned filament is dominated by approaching galaxies
with respect to the RG, suggesting that those galaxies are falling onto the protocluster 
core along the filament and penetrating into the very centre.
On the other hand, the north-west and east areas are preferentially occupied by receding
populations.
The south-west complex further away from the protocluster centre 
near the bottom right corner of the figure consists solely of approaching galaxies,
suggesting the existence of a coherently moving group or filament.

All these spatial and kinematic structures seem to suggest that the inner part
of the cluster centred on the RG ($<$0.5Mpc) is already collapsed
and nearly virialized, while the outer regions are still highly
structured and are at the early phase of assembly towards the protocluster core.
The radio galaxy PKS1138 is right at the junction of the infalling filaments at the
dynamical centre. 

Upon the assumption of virialization in the core, the dynamical mass of the core
is estimated to 1.71$\times$10$^{14}$ M$_\odot$ from the velocity
dispersion of 683 km s$^{-1}$ within C1.
The X-ray observation of this system with the High Resolution Imager on {\it ROSAT} 
show that the emitted energy is 6.7$\pm$1.3$\times$10$^{44}$ ergs s$^{-1}$ in 2--10 keV band
corresponding to the dynamical mass of $\sim$10$^{14}$ M$_\odot$.
It is consistent with our result, although the X-ray emission is 
contaminated by active galactic nuclei \citep{Carilli:1998, Pentericci:2002}. 

\subsection{USS1558-003 ($\mathbf{z}$=2.53)}

USS1558 is the densest protocluster ever known to date at $z$$>$2, and was 
first discovered by \citet{Kajisawa:2006} and H12. 
H12 identified three groups of HAEs aligned along the north-east--south-west directions: a loose group around the RG,
the richest clump at 3.5 arcmin away from the RG, and a small clump in between the two.
This work revises that previous grouping to two clumps C1 and C2 as shown in Fig.\ \ref{fig2}(b)
because of the poor kinematical separation between the latter two clumps, identified only
spatially in the previous study.
We have now confirmed spectroscopically that the two groups are actually
located at the redshift of the RG and are embedded in the large-scale structure.  
C2 is particularly rich, as it confines 19 spectroscopically confirmed HAEs
and 12 more candidates (not confirmed yet) within radius 0.6 Mpc and 
is the densest system ever identified at high redshifts ($z$$>$2).
We note that the RG itself seems to be offset from this densest clump,
unlike the PKS1138 cluster.

Judging from the very high densities of the HAEs in compact areas,
we could reasonably assume a local virialization in these regions, 
and the corresponding dynamical masses and R$_{200}$ are estimated and listed in Table \ref{tab2}.
The dynamical mass of C2 is estimated to
0.87$\times$10$^{14}$ M$_\odot$ from its velocity dispersion of 574 km s$^{-1}$.

We find that there is a large-scale velocity gradient across the cluster
in the direction of the group alignment (north-east--south-west).
The velocity distribution and central value of the south-west group (C2) are blue-shifted 
from those of the north-east group (C1) that hosts the RG (Fig.\ \ref{fig2}; see also Table \ref{tab2}).
Therefore those groups are probably physically aligned and gravitationally pulling each other closer.
They would eventually merge together and become a single rich cluster in the near future. 

\begin{figure}
\centering
\includegraphics[width=80mm]{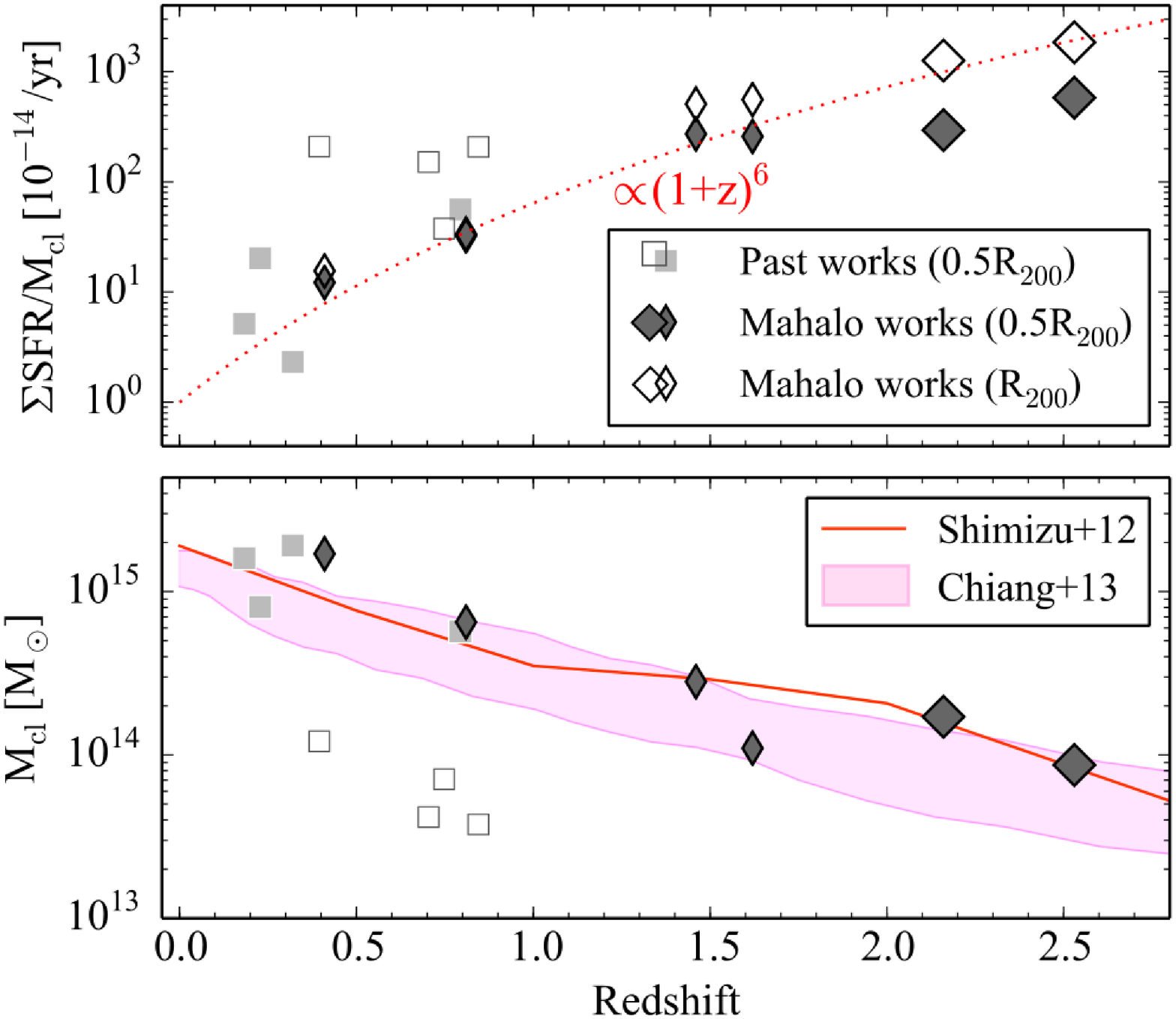}
\caption{(a) Integrated SFRs in the cluster cores, $\Sigma$SFR,
normalized with the cluster dynamical masses (upper panel) and (b) cluster dynamical masses 
(lower panel) are plotted as a function of redshift.
The open diamonds show the measurements for the Mahalo-Subaru cluster sample including this work, 
calculated within $R_{200}$ of each cluster \citep{Koyama:2010, Koyama:2011, Hayashi:2011, Tadaki:2012}. 
The filled diamonds indicate the values within 0.5$R_{200}$ to match the definition with
other previous measurements for a direct comparison.
The previous works for eight clusters at $z$=0.1--0.9 are shown by squares (compilation by \citealt{Finn:2005}). 
The grey and open squares separate those clusters according to their dynamical masses, 
as shown in the lower panel. 
Note that SFRs are measured from \ha\ line strengths for all the clusters except for
the $z$=1.46 and $z$=1.62 clusters which are based on \oii\ lines.
Note also that $R_{200}$ is not fully covered for the $z$=0.81 cluster.
The cluster mass of J0218 cluster at $z$=1.62 is adopted from \citet{Tanaka:2010}.
The red dotted curve shows the relation as a function of redshift, scaling as (1$+$$z$)$^6$.
In the lower panel, the red line and the pink zone show the typical mass growth history of massive
cluster haloes with 1--2$\times$10$^{15}$\msun\ predicted by theoretical models \citep{Shimizu:2012, Chiang:2013}. 
The pink zone corresponds to $\pm$1$\sigma$ scatter around the median values. 
}
\label{fig3}
\end{figure}

\subsection{Cosmic evolution of $\mathbf{\Sigma}$SFR/M$_{\rm\mathbf{cl}}$}

Finally, we investigate the cosmic SF history in galaxy clusters represented by the integrated 
SFR ($\Sigma$SFR) normalized by cluster dynamical mass (Fig. \ref{fig3}a). 
In order to compare our results directly with the previous works compiled by \citep{Finn:2005},
we sum up individual SFRs of the \ha\ (or \oii) emitters within 0.5$R_{200}$,
including the candidates whose membership has not been confirmed yet.
In this work, we calculate $\Sigma$SFR and M$_{cl}$ for the main body of the PKS1138 cluster (C1)
and the richest clump of USS1558 cluster (C2), respectively.
We assume an uniform dust extinction of A$_{H\alpha}$=1 and A$_{\rm\,[O{\sc II}]}$=1.76A$_{H\alpha}$.
To evaluate the uncertainties from this assumption, we also estimate $\Sigma$SFR within R$_{200}$, 
employing the mass-dependent correction for dust extinction \citep{Garn:2010}.
Although the absolute values of $\Sigma$SFR may have large systematic errors, due to various factors
such as sampling bias, active galactic nucleus contribution and so on, the relative differences among 
different clusters that we see here are more reliable.
\citet{Koyama:2010, Koyama:2011} found that $\Sigma$ SFR/M$_{cl}$ in cluster cores 
increases dramatically to $z$$\sim$1.5 and scales as $\sim$(1+$z$)$^6$ (see also \citealt{Smail:2014}). 
We find that this trend extends to even higher redshifts to $z$$\sim$2.5, as
our values of $\Sigma$SFR/M$_{cl}$ within $R_{200}$ estimated in this work 
seem to more or less follow the extrapolated curve of the redshift evolution.

In such comparison of clusters at different redshifts, we must be sure that we are comparing
the right ancestors with the right descendants. 
In fact, clusters grow in mass with cosmic times by a large factor and therefore we should compare 
galaxy clusters taking into account such mass growth.
Figure \ref{fig3}(b) show cluster masses as a function of redshift.
The red line and the pink zone show the mass growth history of massive cluster haloes predicted 
by cosmological simulations \citep{Shimizu:2012, Chiang:2013}.
The data points show the measurements of dynamical masses of real clusters
used for comparison.
It turns out that our protoclusters at $z$$>$2 have large enough masses to be consistent with
the progenitors of the most massive class of clusters, like Coma. 
The lower-$z$ clusters, shown with filled squares also follow the same mass growth curve.
Therefore we argue that we are comparing the right ancestors with right descendants and the
redshift variation of the mass-normalized SFRs seen in the upper panel can be seen
as the intrinsic cosmic SF history of the most massive class of clusters. 

In this Letter, we have presented the kinematical structures of the two richest
protoclusters at $z$$>$2, and extended the cosmic evolution of $\Sigma$SFR/M$_{cl}$ 
back to $z$$>$2 or 11 Gyrs ago, based on the intensive multi-object NIR spectroscopy of the 
NB-selected SF galaxies. 
In our forthcoming Paper I\hspace{-.1em}I (in preparation), we will discuss the physical 
properties of these galaxies (such as gaseous metallicities, ionizing states and
dust extinctions) using the multi-emission-line diagnostics, and compare them with those in 
the general field at similar redshifts. 

\section*{Acknowledgments}

We thank Prof. T. Yamada at Tohoku University for allowing us to use the VPH-K grism on MOIRCS. 
We also acknowledge Dr. K. Aoki at Subaru Telescope for useful discussions. 
This work is financially supported in part by Grant-in-Aid for the Scientific Research
(Nos.\, 21340045 and 24244015) by the Japanese Ministry of Education, Culture, Sports,
Science and Technology. 
We are grateful to the anonymous referee for useful comments. 


\bsp

\label{lastpage}

\end{document}